\documentclass[useAMS]{mn2e}
\usepackage{times}
\usepackage{psfig}

\newcommand{\om}{\omega}

\newcommand{\La}{\Lambda}

\newcommand{\f}{\frac}

\def\be{\begin{equation}}
\def\ee{\end{equation}}
\begin{document}
\title[Evidence of Strange
stars revisited .....]{Evidence for strange stars from joint
observation of harmonic absorption bands and of redshift.}
\author[Bagchi et al. ...]{Manjari Bagchi $^{1}$, Subharthi Ray $^{2}$, Mira Dey
$^{1,~*}$ and   Jishnu Dey $^{1}$\\
$^1$ Dept. of Physics, Presidency College, 86/1 , College Street,
Kolkata 700 073, India. \\$^2$ Inter University Centre for Astronomy and
Astrophysics,
Ganeshkhind, Pune 411 007, India.
\\ $^*$  CSIR Emeritus Professor.}

\maketitle

\begin{abstract}
{From recent reports on terrestrial heavy ion collision
experiments it appears that one may not obtain information about
the existence of asymptotic freedom (AF) and chiral symmetry
restoration (CSR) for quarks of QCD at high density. This
information may still be obtained from compact stars - if they are
made up of strange quark matter (SQM).\\ Very high gravitational
redshift lines (GRL), seen from some compact stars, seem to
suggest high ratios of mass and radius ($M/R$) for them. This is
suggestive of strange stars (SS) and can in fact be fitted very
well with SQM equation of state (EOS) deduced with built in AF and
CSR. In some other stars broad absorption bands (BAB) appear at
about $\sim 0.3~keV$ and multiples thereof, that may fit in very
well with resonance with harmonic compressional breathing mode
frequencies of these SS. Emission at these frequencies are also
observed in six stars.\\ If these two features of large GRL and
BAB were observed together in a single star, it would strengthen
the possibility for the existence of SS in nature and would
vindicate the current dogma of AF and CSR that we believe in QCD.
Recently, in 4U 1700$-$24, both features appear to be detected,
which may well be interpreted as observation of SS - although the
group that analyzed the data did not observe this possibility. We
predict that if the shifted lines, that has been observed, are
from neon with GRL shift $z~=~0.4$ - then the compact object
emitting it is a  SS of mass 1.2 $M_\odot$ and radius 7 km. In
addition the fit to the spectrum leaves a residual with broad dips
at 0.35 $keV$ and multiples thereof, as in 1E1207$-$5209 which is
again suggestive of SS. }
\end{abstract}


\section{Introduction}

QCD, the theory of strong interactions has some peculiar built-in
features which should be tested. For example if particles are
densely packed, they are believed to be free in the theory due to
the property AF. Again the most important building blocks of the
theory, namely (u,d,s) quarks, are believed to be nearly massless
but acquires a substantial mass when the density is low. The
masses are supposed to become small when the density is high
(CSR). The third surprise in QCD is that quarks are completely
confined in a hadron and one can only question whether in a large
system like a compact star there may be partial deconfinement in
the sense that quarks are not bound as in a hadron. The study of
SS from theory and observations is thus a very interesting
challenge - Itoh thought of quarks with parastatistics as
constituents of stars as early as 1970, even before QCD. Then
Witten (1984) discussed the existence of strange quark matter
(SQM) and strange stars (SS) as a result of cosmic separation of
phases.

The possible existence of SQM, with AF and CSR, is studied in
astrophysics as there are (a) some evidence of compact objects
that fit in with SS (Dey et al. 1998, Li et al. 1999a, Li et al.
1999b) and (b) some more evidence from two quasi-periodic peaks in
the power spectrum of accreting low mass X-ray binaries (Li et al.
1999b and Mukhopadhyay et al. 2003). Many superbursts are now
seen. They are like normal Type I thermonuclear bursts and seen
from some compact stars which accrete from their binary partners -
but last for hours instead of seconds and are highly energetic.
One may need to invoke SS to explain them (Sinha et al. 2002,
Sinha et al. 2005, Page \& Cumming 2005). A limiting value of the
magnetic field can be assigned to strange stars of $\sim 10^8 ~G$
explaining he shortage of radio pulsars with fields less than this
value (Mandal et al. 2006).

The situation is worse in heavy ion reactions for testing the
features of QCD. The recent reports of the gold on gold collisions
from four groups BRAHMS, PHENIX, PHOBOS and STAR (Arsene et al.
2004, Adler et al. 2005, Black et al. 2005, Adams et al. 2003)
indicate a phase transition from ordinary matter. But this phase
is not the quark gluon plasma with AF and CSR that the
protagonists searched for :

 ``However, there are a number of features, early on considered as
defining the concept of the QGP, that do not appear to be realized
in the current reactions, or at least have not (yet?) been
identified in experiment. These are associated with the
expectations that a QGP would be characterized by a vanishing
interaction between quarks and exhibit the features of chiral
symmetry restoration and, furthermore, that the system would
exhibit a clear phase transition behaviour." (BRAHMS, Arsene $et.
~al$ 2004)

The strange star hypothesis  may thus be a better laboratory to
test the properties of QCD.

\section{The Model}

't Hooft (1974) suggested the use of inverse of the number of
{\it{colours}} $1/N_C$ as an expansion parameter since there are
no free parameters in QCD. A baryonic system like stars can be
explored in relativistic tree level calculation with a potential
which may be obtained from the meson sector phenomenology in
(Witten 1979) using $1/N_C$ theory. We have done this with
Richardson potential - modified to have different scales for AF
and confinement - deconfinement mechanism (CDM) with parameters
are fitted from baryon magnetic moments (Bagchi $et. ~al$ 2005).
So there are no free potential parameters left in SS calculations.
In this model, a density dependent mass is chosen and from the two
body potential a mean field EOS is derived for beta stable
chargeless (u-d-s) matter. The beta stability and charge
neutrality demands a self consistent calculation of the chemical
potentials of the quarks and electrons since interquark
interaction is present and also involves contribution from density
dependence of quark mass and gluon screening length.

Our interaction potential is due to Richardson (1979) and is given
by,
 \be V_{ij} = \f{12
\pi}{27}\f{1}{ln(1 + {\bf q}^2 /\La ^2)}\f{1}{{\bf q}^2}
\label{eq:V} \ee ${\bf q}\equiv{\bf k}_i - {\bf k}_j$ being the
momentum transfer between i$^{th}$ and j$^{th}$ quarks. The scale
parameter ${\La}$ is $\sim400~MeV$ for hadron phenomenology
(Crater \& van Alstine 1984, Dey $et.~al$ 1986). But in SS
calculation ${\La}$ is taken $\sim100~MeV$ (Dey $et.~al$ 1998).
This discrepancy ultimately leads us to modify the potential and
the modified form of the potential is,
\begin{eqnarray}
\nonumber V_{ij} &=& \frac{12 \pi}{27}\left[(\frac{1}{{\bf
q}^2{\rm ln}(1 + \frac{{\bf q}^2}{\Lambda
^2})}-\frac{\Lambda^2}{{\bf q}^4}) +\frac{{\Lambda^\prime}^2}{{\bf
q}^4}\right] \label{eq:Vtl}
\end{eqnarray}

${\Lambda}$ is the scale parameter representing the AF as the
first two terms are asymptotically zero for large ${\bf q}$.
${\Lambda}^{ \prime}$ is the scale parameter corresponding to
confinement as the third term reduces to a linear confinement for
small ${\bf q}$. From Bagchi $et.~ al$, (2004), we get
${\Lambda}~=~100 ~MeV$ and ${\Lambda}^{\prime}~=~350 ~MeV$. The
modified potential is used in SS calculation (Bagchi $et.~al$,
2005b). The potential is screened due to gluon propagation in a
medium and ${\bf q}^2$ is replaced by $[{\bf q}^2+D^{-2}]$.
Inverse screening length $D^{-1}$ ($i.e.$ gluon mass $m_g$) is :
\begin{equation} (D^{-1})^2 ~=~ \frac{2 \alpha_0}{\pi}
\sum_{i=u,d,s,}k^f_i \sqrt{(k^f_i)^2 + M_i^2} \label{eq:gm}
\end{equation} Fermi momenta are related to number density:
\begin{equation}
{k^f_i} = (n _i \pi^2)^{1/3} \label{eq:kfq}
\end{equation}
\begin{equation}
{k^f_e} = (n _e 3 \pi^2)^{1/3} \label{eq:kfe}
\end{equation}
The subscript $e$ stands for the electron. Perturbative quark
gluon coupling, $\alpha_0$ is $0.2$ in the unmodified potential
and 0.55-0.65 for the modified potential. The quark mass, $M_i$ is
taken to be density dependent to restore chiral symmetry at high
density :
 \be M_i = m_i + M_Q ~~sech\left( \f{n_B}{N n _0}\right),
\;\;~~~ i = u, d, s. \label{eq:qm} \ee where the constituent
quark mass $M_Q$ lies between 300-350 $MeV$ for each quark, $n_B
= (n _u+n _d+n _s)/3$ is the baryon number density, $n _0$ is the
normal nuclear matter density and $N$ is a parameter. The current
quark masses, $m_i$ are taken as 4, 7 and 150 $MeV$ for u, d, s
quarks respectively. From the charge neutrality and beta
equilibrium conditions, $k_i^f$ and $\mu_i$ (chemical potentials)
of quarks and electron are obtained as a function of $n_B$ and
performing a relativistic Hartree-Fock calculation, the energy
density $\epsilon$ of the SQM is obtained. The first law of
thermodynamics gives the EOS of SQM at zero temperature as:
\be
P = \sum_i({\mu_i}{n_i}-{\epsilon_i}) +
({\mu_e}{n_e}-{\epsilon_e}) \label{eq:eos} \ee To obtain a
realistic EOS, the model parameters $\alpha_0$, $M_Q$ and $N$ are
chosen in such a way that the minimum value of energy per baryon
($E/A~\equiv~{\epsilon}/{n_B}$) for uds quark matter is less than
that of the most stable element $Fe^{56}$ $i.e.$ 930 $MeV$. Thus
uds quark matter can construct stable stars. The minimum value of
$E/A$ is obtained at the star surface where the pressure is zero.
The presence of zero pressure indicates the existence of a sharp
surface of the strange star in contrast to fuzzy surfaces of the
neutron stars. The surface is sharp since strong interaction
dictates the deconfinement point.

On the contrary, for the same parameters, the minimum value of
$E/A$ for ud quark matter is greater than that of $Fe^{56}$ so
that $Fe^{56}$ remains the most stable element in the non-strange
world. Using this EOS for strange quark matter we obtain the
Mass-Radius relation of the strange star by solving the
hydrostatic equilibrium equation (TOV equation) with appropriate
boundary conditions.

Among different EOSs obtained from our SS model, varying the
parameters, only one (EOS A) is used in the present work and the
parameters for this EOS are listed in table \ref{tab:allmrnew}.
The variation for other parameter sets are unimportant.

\begin{table*}
\begin{center}
\caption{EOS A using the modified potential. ${\Lambda}$ is 100
$MeV$. ${\epsilon}_c/c^2$ is the central density.} \vskip 0.5 cm
\label{tab:allmrnew}
\begin{tabular}{|c|c|c|c|c|c|c|c|c|c|}
\hline
EOS&${\Lambda}^{\prime}$&$M_q$&$N$&${\alpha_0}$&${(E/A)}^{uds}_{min}$
&${(E/A)}^{ud}_{min}$ &$M_{max}$& R & ${\epsilon}_c/c^2$\\
Label  &  &  & &&  & &  &  for $M_{max}$ & for $M_{max}$ \\
 &  &  & &&  & &  &  & $10^{14}$\\
&$MeV$&$MeV$ & & &$MeV$& $MeV$&$M_\odot$& $km$& $gm~cm^{-3}$\\
\hline \hline

A& 350&325 &3.0 &.55 &874 &942& 1.53 &7.41 & 42.98 \\

\hline \hline
\end{tabular}
\end{center}
\end{table*}

\section{Harmonic bands for absorption.}

In our SS model, the energy per baryon has a minimum at the star
surface and this surface can vibrate
(Sinha et al., 2003 and Ray et al., 2004). A Taylor expansion of
the energy per baryon about $r=R$ can be written as
\be
E(r)/A =
E(R)/A + \f12k(R)~~ (r-R)^2
\ee
which gives
$\left(\f{dE}{dr}\right)_{r=R}=0$ and
$k(R)~=~\left(\f{d^2E}{dr^2}\right)_{r=R} $. Here r is the radial
coordinate, R is the radius of the star and $E(r)/A$ is the
energy per baryon at $r$. The fundamental frequency of the
vibration is given by :
\be
\om = \sqrt\f{k(R)}{m_{skin}}
\ee
and
the energy of the vibrational mode as $\hbar\om$. This
fundamental mode and its harmonics then should be observable in
the spectrum of the star.

To calculate $\om$, pressure (p) is written as :
\be
p = -\f{dE}{dV}~=~-\f1{4\pi r^2}~\f{dE}{dr} \ee using the first
law of thermodynamics for zero temperature system. Then $k(R)$ can
be written as
\be
k(R) = -4\pi~R^2\left(\f{dp}{dr}\right)_{r=R}
\label{kR}
\ee
$\left(\f{dp}{dr}\right)_{r=R}$ is obtained by solving the TOV
equation as mentioned in the previous section. The mass of the
skin $m_{skin}$ is calculated from the properties of electron
cloud around the star (see Sinha et al. 2003 for details).

The existence of absorption bands at 0.35 $keV$ with harmonics at
0.7, 1.05, 1.4 etc, in 1E 1207$-$5209 were interpreted in Sinha et
al. (2003) as evidence of surface vibration from strange stars.
This is an isolated star and presumably there is no accretion disk
associated with this star.

It was then observed that six stars accreting stars show enhanced
harmonic band emission at energies corresponding to the same
vibrational modes (Ray et al. 2004). More evidences of such kind
of surface vibration will help to establish strange star model.
These stars are accreting and thus there is a disk above it. In
SS, the electrons tend to flow out from the strongly held quarks
since they do not feel the QCD forces. Only the net resulting
balancing positive charge of the star holds the electrons close to
the star - they form an envelope of about few hundred Fermi above
the star as first shown by Alcock, Farhi and Olinto (1986).

The electron cloud outside and the positive central charge leads
to a electrostatic field which may be as strong as
$5\times10^{17}~ {Vcm^{-1}}$ decreasing to $10^{11}~Vcm^{-1}$ as
one goes out radially by $10^{-8}~ cm$ (Xu, Zhang and Qiao 2001).
The accretion disk on top of the electron layer gathered by the
star from their binary partner lies typically within the range
$10^{-11}M_{\odot}~yr^{-1}< \dot{M}< 10^{-8}M_{\odot}~yr^{-1}$
(Cumming and Bildsten, 2000).

There is a coupling of the surface vibration of the SS and the
accretion disk that leads to the enhancement in select harmonic
bands of the emission spectrum. We hope more data will become
available in near future.

\subsection{4U~1700$+$24}
\begin{figure}
\centerline{\psfig{figure=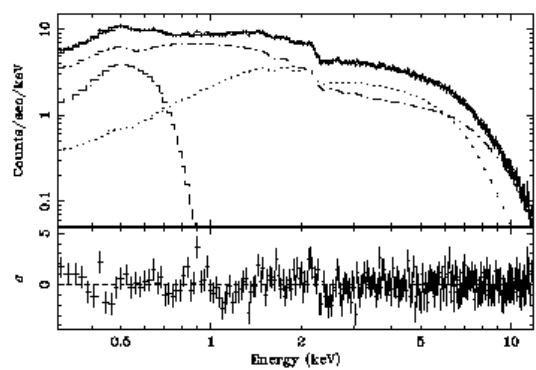,width=8cm}} \caption{The
PN count rate spectrum of 4U 1700$+$24 and residuals in units of
$\sigma$ with respect to the best fit model. The individual model
components are also shown. The presence of absorption lines is
clear from the residuals. This graph is taken from Tiengo et al.
(2005).} \label{fig:tiengo}
\end{figure}

In Fig.(\ref {fig:tiengo}) we display the curve given in Tiengo et
al. (2005) for the spectral analysis of 4U 1700+24 for which
possibly a line with redshift $z\sim0.4$ was observed (will be
discussed in the next section). Note the residual at the bottom
where there are the absorption bands visible upto 2 $keV$. We hope
in future, when much better data will become available, one will
be able to test our conjecture that these bands are also due to
surface vibration as the emission bands in six stars and the
absorption bands in 1E1207$-$5209 discussed above.

\begin{table}
\caption{Coefficients of the Gaussian used to fit the data given
by Mereghetti.} \vskip .5cm

\begin{center}
\begin{tabular}{|c|c|c|c|c|c|c|c|c|c|}
\hline

$\epsilon$'s &0.36 &0.72 & 1.08&1.44&1.80&2.16 & 2.52&2.88
\\ \hline
 $a$'s &2.8 &10.0 &1.5 &6.0 &1.0 &2.0 &3.0 & 1.0  \\ \hline
$\sigma$'s &800 &35 &110 &55 &700 &90 &10 &70   \\ \hline
\end{tabular}
\end{center}
\label{tab:coef}
\end{table}

\begin{figure}
\centerline{\psfig{figure=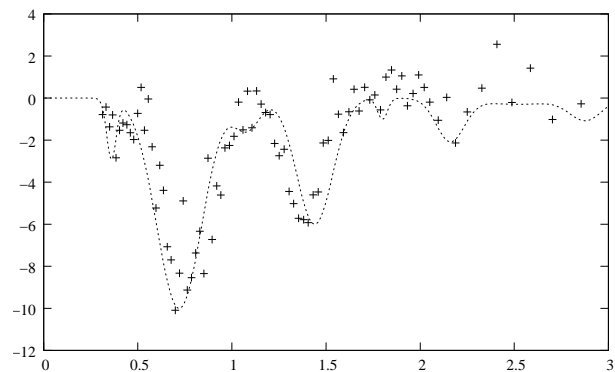,width=8cm}} \caption{The fit
of the residual provided by Mereghetti with a Gaussian.}
\label{fig:mere}
\end{figure}

\subsection{1E1207$-$5209 Revisited.}
As already mentioned, the absorption bands at 0.35 $keV$ with
harmonics at 0.7, 1.05, 1.4 etc, in 1E 1207$-$5209 were
interpreted in Sinha et al. (2003) as evidence of surface
vibration from strange stars.

Now, the residuals for 1E1207$-$5209, as supplied by S.
Mereghetti, is plotted in Fig.(\ref{fig:mere}) and are fitted
with an empirical form for harmonic oscillations with an arbitrary
envelope of the form
\begin{equation}
\chi(E) = \sum_{j=1,8}a_j exp[-{\sigma}_j(E
-{{\epsilon}_j})^2]
\end{equation}
where $a_j$s are the amplitudes and
$\epsilon_j$s are the central frequencies with widths
$\sqrt{2}/{\sigma_j}$. $a_j$ and ${\sigma}_j$ are given in the
table \ref{tab:coef}. It is clear that the fundamental and even
harmonics are suppressed while the odd harmonics are dominant.
This may be due to the preferential detection efficiency of the
signal receiver, in which case the observed form will change when
a different satellite is used. On the other hand if it is due to
preferential absorption in the interstellar medium, the form
could change even with the same detector.

To us the second alternative seems preferred at the moment, since
at different times with change of phase, shown in different colours in
Bignami et al. (2004). The intensity in these
phases vary. But a clear understanding of the physical phenomena
behind it is yet needed.

\section{The redshift for various star mass and radius.}

Another check for the existence of SS is the mass-radius ($M/R$)
ratio which can be inferred if the gravitational redshift is
measured for some spectral lines. The gravitational redshift is
given by :

\begin{equation}
z = \left(1-\frac{2GM}{c^2R}\right)^{-1/2} - 1
\end{equation}

\begin{figure}
\centerline{\psfig{figure=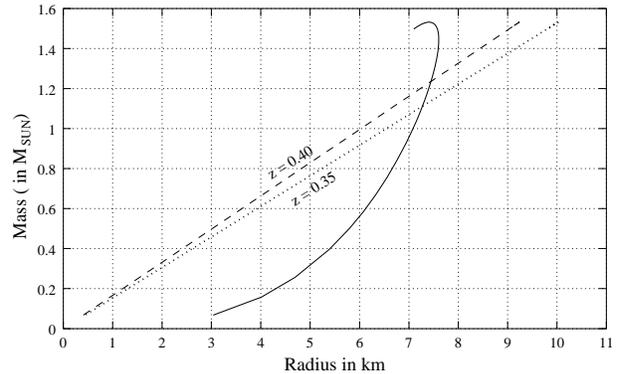,width=8cm}} \caption{The
mass radius curve for ReSS (EOS A of Bagchi et al. 2005) showing
the intersections for $z$ values of 0.35 found by Cottam et al.
(2002) and 0.4 implied in Tiengo et al. (2005)  if the lines
around 19 \AA $~$are due to Neon in the accretion disk.}
\label{fig:mrz35_40g}
\end{figure}

From the spectral analysis of X-ray bursts from EXO 0748$-$676
Cottam, Paerels and Mendez (2002) found a redshift of $z~=~0.35$.
Later, Tiengo $et.~al$ (2005) performed a spectral analysis of an
XMM-Newton observation of the X-ray binary 4U 1700+24 during an
outburst in August 2002. At low energies they detected a
prominent soft excess, which they modeled with a broad Gaussian
centered at $\sim$ 0.5 keV. In the high resolution RGS spectrum
they detected an emission line, centered at
$19.19^{+0.05}_{-0.09}~$\AA$~$ . The authors gave two possible
interpretations for this line: O VIII at redshift $z \sim 0.012$
or Ne IX at redshift z $\sim$ 0.4. But the authors argued O VIII
to be a better candidate than Ne for the observed emission line.
On the contrary, while modeling the spectrum of three LMXBs 4U
0614$+$091, 2S 0918$-$549, and 4U 1543$-$624 Juett et al. (2003)
proposed that there is excess neon local to each of these sources
as in the 4U 1626$-$67. So there is no reason to exclude the
possibility of 19.2 \AA $~$line from 4U 1700+24 to be a Ne one
giving $z \sim 0.4$ and the star to be a strange star. In
Fig.(\ref{fig:mrz35_40g}), we plotted two lines one for $z=0.35$
and the other for $z=0.40$ in the Mass(M)-Radius(R) parameter
space which cuts the M-R relation obtained form our EOS giving
realistic values of mass and radius. For $z~=~0.35$ we find a
star of mass slightly larger than 1.1 $M_\odot$ (fulfilling the
condition M $\geq$ 1.1 for EXO 0748$-$676 as demanded by Cottam,
Paerels and Mendez (2002)) and radius 7.0 $km$. Again, for 0.4 we
get a star with mass 1.2 $M_\odot$ and radius 7.5 $km$. So the
possibility of EXO 0748$-$676 and 4U 1700+24 to be strange stars
are quite open.

On the other hand, Sanwal $et.~al$ (2002) discussed that the
absorption lines from 1E1207$-$5209 have $z~=~0.12-0.23$ which
corresponds to M $~0.36-0.76~M_{\odot}$ and R $~5.19-6.56~km$.
These values may appear to be unrealistic, but the fact is that
the authors explained the absorption lines as a result of atomic
transitions while in SS model the origin of absorption lines are
completely different (as discussed in the previous section).

Therefore, contradicting current quotes in the literature (Tiengo
et al. 2005, Cottam et al. 2002 and Miller 2002), our SS model can
explain redshifts observed from some compact stars.

In the next figure we plot with mass and radius of the strange
star and see that for $z~=~0.6$ one can get a star of mass 1.55
$M_\odot$ and the radius is somewhat smaller - about 7.4 $km$. We
hope that a look at these curves will dispel the mistaken notion
(Cottam et al. 2002 \& Miller, 2002) that strange stars cannot
fit EXO 748$-$676.

\begin{figure}
\centerline{\psfig{figure=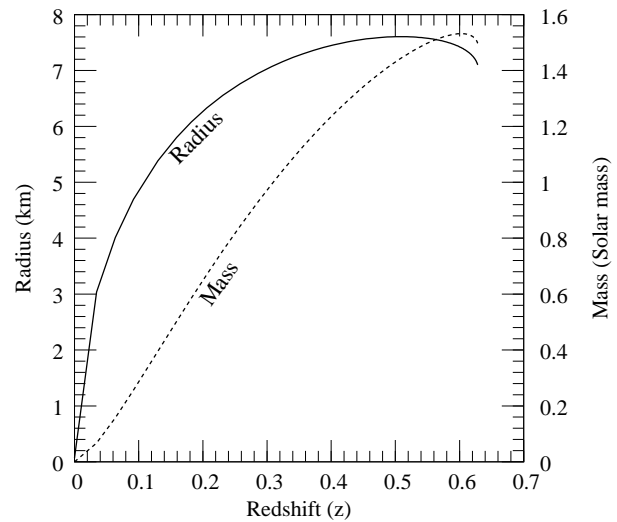,width=8cm}} \caption{The
plot of mass and radius vs redshift, where the masses for the SS
are taken for EOS A given in Bagchi et al. (2005). This curve
shows clearly that for observed redshift 0.3 to 0.6, the SS may
have mass between 1.0 to 1.5 $M_{\odot}$ and radius between 7.1
to 7.6 km. } \label{fig:mrz}
\end{figure}

\section{Discussion}

We now discuss the possibility of anti strange stars discussed by
Boehm et al. (2003) and Oaknin \& Zhitnitsky (2005). If such stars
exist then they will have an atmosphere of positrons rather than
electrons spilling outside the star radius by a few hundred $fm$.
These positrons will then interact with electrons of the normal
matter accreting onto the strange star from its binary partner and
positron-electron annihilation will produce the 511 $keV$
$\gamma$-pair. The only snag in this scenario was that for these
anti-quark stars to be stable the surface tension has to be large
(Alcock \& Olinto, 1989). Recently we have shown that in our ReSS
model the surface tension is large unlike the bag model (Bagchi et
al. 2005b) and such anti-strange stars are indeed likely to be
stable. This partially solves the problem of baryonic dark matter
and baryon asymmetry of the universe.

One must affirm that it is quite possible that there are two types
of compact stars, namely neutron stars (NS) and strange stars.
Indeed, a clear possibility of existence of these two kinds of
stars  is discussed very recently by Bombaci, Parenti and Vidana
(2004) and these authors showed a phase transition from NS to SS
can account for delayed gamma ray bursts.

\section{Summary and conclusion}

We have shown that the observed gravitational redshift $z$ can be
explained by our SS model with built in properties of QCD. Also
harmonic bands in the spectrum may convincingly prove the
existence of strange stars.

We hope that with the profuse data that is flowing in, one will
find more stars like 4U 1700$+$24 so that the strange star
hypothesis can be put through a stringent test. Further data on
residuals for 1E1207$-$5209 will also be very interesting.

\section*{Acknowledgments}
We thank Dr. Sandro Mereghetti for promptly sending us the data
used in Fig.(\ref{fig:mere}) by email and Prof. van der Klis for
permission to use their figure for $4U 1700+24$ (Fig.\ref{fig:tiengo}). MB, MD
and JD acknowledges HRI, Allahabad, India, and IUCAA, Pune, India, for fruitful
visits. They also acknowledges DST grant no. SP/S2/K-03/2001,
Govt. of India, for research support.


\end{document}